\documentstyle[aps,prb]{revtex}
\newcommand{\eqref}[1]{(\ref{#1})}

\begin{document}

\title{\bf{A Study of Quasi-Exactly Solvable Models within the Quantum
Hamilton-Jacobi Formalism }}
\author{K. G. Geojo\thanks{akksprs@uohyd.ernet.in}, S. Sree
Ranjani\thanks{akksprs@uohyd.ernet.in}, A. K.
Kapoor\thanks{akksp@uohyd.ernet.in}\\
\address{School of Physics, University of Hyderabad,\\Hyderabad 500 046,
INDIA}}

\maketitle

\begin{abstract}
A few quasi-exactly solvable models are studied within the quantum
Hamilton-Jacobi formalism. By assuming a simple singularity structure of
the quantum momentum function, we show that the exact quantization
condition leads to the condition for  quasi-exact solvability.
\end{abstract}

%\maketitle
%\newpage

\section{Introduction}
This paper reports an investigation of quasi-exactly solvable (QES) models
within the quantum Hamilton-Jacobi (QHJ) formalism. One of the earliest
investigations on QES models was done by V. Singh {\it et.al} [1] . The
QES models have been studied extensively and for
a general review and references we refer the reader to the book by
Ushveridze [2]. In the recent years, a lot
of interesting work has been done on QES periodic potentials. We refer the
reader to  references [3-13]. A complete list of canonical forms for the
QES models has been 
obtained in [3,4]. The QES models have a characteristic property that,
when the
potential parameters satisfy a specified condition, analytic expressions
for a few energy levels
and their corresponding eigenfunctions can be obtained exactly. This
condition
between the parameters of the potential will be referred to as
quasi-exactly
solvability condition.

A well known example of QES model is the sextic oscillator in one
dimension,
corresponding to the potential $ V(x) =\alpha x^{2}+\beta x^{4}+\gamma
x^{6}$. The condition of quasi-exactly solvability is found to be
$\frac{1}{\sqrt{\gamma}}[\frac{\beta^{2}}{4\gamma}-\alpha] = 3+2n$, where
$n$
is a non-negative integer which is related to the number of levels for
which
exact energy eigenfunctions and eigenvalues can be computed. A number of
other
QES models have been constructed and studied within algebraic and group
theoretical approach. Interesting connections have been established
between
the sextic oscillator and second order linear differential equations
within a
new approach to second order linear differential equations [13].

In this paper we study the sextic oscillator and a few other QES
models in one dimension. In each of these cases the condition of
quasi-exactly solvability is derived within the QHJ approach. The next
section contains an overview of QHJ formalism and how it is used in
the present paper to study the QES models. In the following four
sections we investigate the sextic oscillator, sextic oscillator with a
centrifugal barrier, a hyperbolic potential and a circular
potential. The last section contains our conclusions about exact
solvability and quasi-exactly solvability.

\section{Quantum Hamilton-Jacobi Formalism}

The QHJ formalism was initiated by Leacock and Padgett [14,15] and was
successfully applied to
several exactly solvable models (ESM) in one dimension in quantum
mechanics by Bhalla {\it et.al} [16-18]. Our discussions in this paper
will be limited to the  QES problems in one dimension only.
The Schr$\ddot{o}$dinger equation is
\begin{equation}
-\frac{\hbar^{2}}{2m}\frac{d^{2}}{dx^{2}}\psi + V(x)\psi = E\psi.
\label{e1}
\end{equation}
On substituting $\psi = \exp({iS/\hbar})$, in \eqref{e1}{} we obtain the
following equation for $S$
\begin{equation}
\left(\frac{dS}{dx}\right)^2 - i \hbar\left(\frac{d^{2}S}{dx^{2}}\right) =
2m[E-V(x)].   \label{e2}
\end{equation}
We define $p(x) = \frac{dS}{dx}$ which 
satisfies the Riccati equation
\begin{equation}
p^{2}(x) - i\hbar\frac{d}{dx}p(x) = 2m[E - V(x)].    \label{e3}
\end{equation}
In the limit $\hbar\rightarrow 0 $, \eqref{e2} becomes the classical
Hamilton-Jacobi
equation and $p(x)\rightarrow  p_{cl}$, the classical momentum which is
\begin{equation}
p_{cl} = \sqrt{2m[E-V(x)]}.    \label{e4}
\end{equation}
Therefore \eqref{e2}{} and \eqref{e3}{} will be referred to as QHJ
equations, $p(x)$ will be called the quantum momentum function (QMF) and
$S$ will be the quantum action. In terms of the eigenfunction $\psi$ of
the energy, the QMF is given by
\begin{equation}
p(x) = -i
 \hbar\frac{1}{\psi}\frac{d\psi}{dx}.  \label{e5}
\end{equation}
An important step in the QHJ formalism is to regard $x$ as a complex
variable and
to extend the definition of $p(x)$ to the complex plane. From \eqref{e5}{}
it is obvious that the zeros of the wavefunction correspond to the poles
of the
QMF. It is known that the bound state solutions of \eqref{e1}{}
corresponding
to the $n^{th}$ level has $n$ real zeros correspondingly the QMF has $n$
poles on the real
line. From \eqref{e3}{} it can be seen that if $x$ is a point at which
$V(x)$ is
analytic and $p$ has a pole, the pole must be of first order and the
residue at that pole will be $- i \hbar$. Therefore the integral of the
QMF
taken along the contour $C$, which encloses these poles, will have the
value
$n\hbar$. Thus we get
\begin{equation}
\frac{1}{2\pi} \oint_{C} p(x)dx = n\hbar.    \label{e6}
\end{equation}
and this is an exact quantization condition in one dimension [14,15].
This exact quantization condition \eqref{e6}{} has been used to obtain
bound
state energy levels without solving for the eigenfunctions for several ESM
[16-18]. For this purpose one needs to know the location of singularities
and
their corresponding residues of the QMF in the complex plane. The QMF has
the
two kinds of singularities, fixed and the moving singularities. The fixed
singularities correspond to the singularities of the potential and will be
present in every solution of the Riccati equation, their location being
independent of the initial conditions. The position of the moving
singularities depends on the initial conditions. It is known that for the
solutions of a Riccati equation only poles 
can appear as moving singularities. Therefore if the potential is
meromorphic, the solutions will
also be meromorphic. Coming back to the QMF corresponding to the solutions
of
the Schr$\ddot{o}$dinger equation, in addition to the poles, corresponding
to
the $n$ real zeros of the $n^{th}$ excited state, in general there may be
other
moving poles. The knowledge about these poles is needed to apply QHJ
method.
The residue at any fixed pole can be computed from \eqref{e3}{}. This,
being a
quadratic equation, will lead to two solutions and therefore a boundary
condition on the QMF is needed to pick up the physical solutions. Leacock
and
Padgett proposed that one should make use of the condition,
\begin{equation}
 \lim_{\hbar \rightarrow 0} p(x)\rightarrow  p_{cl}.  \label{e7}
\end{equation}
From \eqref{e4} it should be noted that the classical turning points
become
the branch points of $p_{cl}$. It has to be emphasized here that, when $x$
is a
complex variable, $p_{cl}$ is a double valued function. A full definition
of
$p_{cl}$ in the complex plane as a function of a complex variable $x$,
requires us to select a particular branch of $p_{cl}$ which is assumed to
have a branch
cut in the classical region and a positive value just below the branch
cut.
Locating the singular points of the QMF and imposing the boundary
conditions
discussed above are two crucial  but usually difficult steps in arriving
at
the correct solutions. In the earlier studies of ESM, it was easy to
guess the singularity structure of the QMF and it was found that,
there were no moving poles away from the real axis. For the QES models
studied in this paper, it is very difficult to find the location of
the singularities of the QMF. In order to be able to make some
progress we make a simplifying assumption about the moving poles
and the nature of the singularity at infinity. The details of moving
and possible fixed poles will be given in each case separately as
and when we discuss the model. We now state our main assumption, common
to all models studied, {\it i.e, the point at infinity is an isolated
point
and that it is a pole of finite order and not an essential singularity}.
Under these assumptions we show that, imposing the exact quantization
condition leads to the condition of QES in each case.

\section{Sextic Oscillator}

The potential for the sextic oscillator is 
\begin{equation}
V(x) = \alpha x^{2} + \beta x^{4} + \gamma x^{6} ,\,\, \gamma > 0.
\label{e8}
\end{equation}
From now onwards we set $\hbar = 2m = 1$.
For the $n^{th}$ exited state, QMF has $n$ moving poles on the real axis
and we
assume that there are no other moving poles. In order to use the
quantization condition \eqref{e6}{} we need to evaluate the integral
$J(E)$
\begin{equation}
 J(E) = \frac{1}{2\pi}\oint_{C} p(x)dx   \label{e9}
\end{equation}
 along the contour $C$. As a first step, it is convenient to deform the
contour 
 $C$ to a large circular contour $\Gamma$, with the center at the origin
and big
enough to enclose all the singularities of the QMF in the finite complex
plane.
This is possible because the point at infinity is assumed to be  an
isolated 
singularity and the value of the integral remains unchanged. Hence
\begin{equation}
J(E) = \frac{1}{2\pi}\oint_{\Gamma} p(x)dx.   \label{e10}
\end{equation}
In order to evaluate the integral along the large circle $\Gamma$, we
apply an inversion mapping $x\rightarrow y = \frac{1}{x}$. Under this
mapping the
circular contour $\Gamma$ gets mapped to another circular contour
$\gamma$  in the $y$ plane. The only singular point inside this contour
is the point $y = 0$, which is the image of the point at infinity in
the $x$-plane. Thus we obtain from \eqref{e10}{}
\begin{equation}
J(E) = \frac{1}{2\pi}\oint_{\gamma}\tilde{p}(y)\frac{1}{y^{2}}dy.
\label{e11}
\end{equation}
To evaluate the integral in \eqref{e11}{} a Laurent expansion of
$\tilde{p}(y)$ is made.
\begin{equation}
 \tilde{p}(y) = \sum_{n=1}^{3}\frac{b_n}{y^n}
 +\sum_{n=1}^{\infty} a_{n}y^{n}.   \label{e12}
\end{equation}
Substituting this expansion in \eqref{e11}{} and integrating term by
term, we get
\begin{equation}
J(E) = i a_1.         \label{e13}
\end{equation}
It only remains to compute the coefficient $a_1$ of the Laurent
expansion given in \eqref{e12}{}. To do this we start from the QHJ
equation
\begin{equation}
\tilde{p}^{2}(y)+iy^{2}
\frac{d}{dy}\tilde{p}(y)=E-\frac{\alpha}{y^2}-\frac{\beta}{y^4}-\frac{\gamma}{y^6}.\label{e14}
\end{equation}
Substituting the Laurent expansion and equating the coefficients of
different powers of $y$ in both sides of the equation \eqref{e14}{}, we
successively get the following equations.
\begin{equation}
b_{3}^{2} = -\gamma.   \label{e15}
\end{equation}
\begin{equation}
b_2 = 0.               \label{e16}
\end{equation}
\begin{equation}
2b_{1}b_{3} = -\beta.  \label{e17}
\end{equation}
\begin{equation}
b_{1}^{2} + b_{3}(2a_{1}-3i) = -\alpha.   \label{e18}
\end{equation}
It is important to know that, we would get two solutions for $b_{1}$
corresponding to the two solutions of $ b_{3}= \pm i\sqrt{
\gamma}$. This happens due to the fact that the QHJ is quadratic in the
QMF. Thus one needs a boundary condition to pick the correct
solution. We propose to use the square integrability of the
wavefunction instead of the original boundary condition, explained
in the introduction, which was proposed by Leacock and Padgett. This is
because the original boundary condition is difficult to implement in
the present case due to the presence of six branch  points in the
$p_{cl}$. In order to find the restrictions coming from the square
integrability, we compute the wavefunction
\begin{equation}
\psi(x) = \exp\left(\int i p(x)dx\right)   \label{e19}
\end{equation}
for large $x$ as follows.
The most important term in the Laurent expansion \eqref{e12}{} for small
$y\approx0$, corresponding to large $x$, is
\begin{equation}
\tilde{p}(y) \approx \frac{b_3}{y^3}   \label{e20}
\end{equation}
and the wavefunction for large $x$ becomes
\begin{equation}
\psi(x) \approx \exp\left(i \frac{b_{3}x^{4}}{4}\right)dx.    \label{e21}
\end{equation}
Out of the two solutions, $b_{3} = \pm i\sqrt{\gamma}$, $\psi$ is square
integrable only for $b_{3} = i\sqrt{ \gamma}$. Using this value of
$b_{3}$ and calculating $J(E)$ from \eqref{e13}{}, the quantization
condition gives

\begin{equation}
\frac{1}{\sqrt{\gamma}}\left(\frac{\beta^{2}}{4\gamma}-\alpha\right) =
3+2n.  \label{e22}
\end{equation}
In order to compare the result in \eqref{e22}{} with the known condition
[2], we write
\begin{eqnarray}
\gamma = a^2  ,\,\,  \beta = 2ab.             \label{e23}
\end{eqnarray}
Thus we get $ \alpha = b^{2} - a(3+2n)$ which agrees with known
result.

\section{Sextic Oscillator with the centrifugal barrier}

This kind of model was first studied by A.P.Houtot [19]. The potential
we discuss here is
\begin{equation}
V(x)=4(S-\frac{1}{4})(S-\frac{3}{4})\frac{1}{x^2}+[b^2-4a(S+\frac{1}{2}+M)]x^2+2abx^4+a^2x^6
\label{e24}
\end{equation}
and the range of $S$ is taken to be $4S>3$. We note that the potential
goes to
$\infty$ as $x \rightarrow 0$ and $x \rightarrow \infty$. Hence the
classical turning points will be on the positive real axis. The
physical motion in both classical and quantum situations will be
confined to positive real axis only. As, has been discussed earlier
there will be n moving poles in the classical region which are
enclosed by the contour $C$ in the quantization condition stated in
\eqref{e6}{}. When we extend the definition of $x$, to take all the
complex values, we expect $n$ additional moving poles on the negative real
axis. These come from the symmetry of the potential under the
transformation $x\rightarrow -x$.
Just as in the quantization condition \eqref{e6}{}, we have
\begin{equation}
\frac{1}{2\pi} \oint_{C_{1}}p(x)dx=n     \label{e25}
\end{equation}
where $C_{1}$ is the contour which encloses the $n$ additional moving
poles on the
negative real axis.
The QHJ equation for this potential is
\begin{equation}
p^{2}(x)-i \frac{d}{dx}p(x) =E-
4(S-\frac{1}{4})(S-\frac{3}{4})\frac{1}{x^2}-[b^2-4a(S+\frac{1}{2}+M)]x^2-2abx^4-a^2x^6.
\label{e26}
\end{equation}
We observe that in addition to the 2$n$ moving poles, $x=0$ is a fixed
pole. In this case we assume that there are no other singularities in the
finite complex plane. As in the case of the sextic oscillator, in
order to evaluate $J(E)$ in \eqref{e9}{}, we deform the contour $C$ to a
large circular contour $\Gamma$, which will enclose the 2$n$ moving
poles and the fixed pole at $x=0$. Therefore we have,
\begin{equation}
\oint_{\Gamma}p(x)dx
=\oint_{\gamma_{0}}p(x)dx+\oint_{C}p(x)dx+\oint_{C_{1}}p(x)dx
  \label{e27}
\end{equation}
where $\gamma_{0}$ is the contour enclosing only the fixed pole at
$x=0$. Expanding $p(x)$ in the Laurent series, in the powers of $x$
\begin{equation}
p(x) = \frac{b_{1}}{x} + \sum_{n=0}^{\infty}a_{n}x^{n} \label{e28}
\end{equation}
we get
\begin{equation}
\frac{1}{2\pi}\oint_{\gamma_{0}}p(x)dx = i b_1.   \label{e29}
\end{equation}
We fix $b_1$ by substituting the Laurent series in
\eqref{e26}{}. Thus we obtain,
\begin{eqnarray}
b_1= \frac{i}{2}[4S-3] ,\,\, b_1= -\frac{i}{2}[4S-1]. \label{e30}
\end{eqnarray}
The choice for the value of $b_1$ consistent with square integrability
in the specified range is
\begin{equation}
b_1 =-\frac{i}{2}[4S-1].   \label{e31}
\end{equation}
Thus \eqref{e27}{} becomes
\begin{equation}
\frac{1}{2\pi}\oint_{\Gamma}p(x)dx = i b_1 + 2n.    \label{e32}
\end{equation}
The integral along $\Gamma$ is computed by changing the variable from
$x\rightarrow y=\frac{1}{x}$ and proceeding in the same way as was
done for the sextic oscillator. From \eqref{e31}{} we then obtain
\begin{equation}
M = n.     \label{e33}
\end{equation}
which is the quasi-exact solvability condition for the sextic
oscillator with the centrifugal barrier.

\section{ Circular Potential}

The potential for this model is
\begin{equation}
V(x)=\frac{A}{\sin^2x}+\frac{B}{\cos^2x}-C\sin^2x+D\sin^4x    \label{e34}
\end{equation}
where,
\begin{equation}
A=4(S_1-\frac{1}{4})(S_{1}-\frac{3}{4})
,\,\,B=4(S_2-\frac{1}{4})(S_{2}-\frac{3}{4})  \label{e35}
\end{equation}
\begin{equation}
C=q_{1}^{2}+4q_1(S_1+S_2+M) ,\,\,D=q_{1}^{2}  \label{e36}
\end{equation}
and the ranges of $S_1$ and $S_2$ are $2S_1>1$ and $2S_2>1$. The QHJ
equation
is
\begin{equation}
p^2(x)-i
\frac{d}{dx}p(x)=E-\frac{A}{\sin^2x}-\frac{B}{\cos^2x}+C\sin^2x-D\sin^4x
\label{e37}
\end{equation}
and the quantization condition is
\begin{equation}
\frac{1}{2\pi}\oint_{C}p(x)dx = n   \label{e38}
\end{equation}
Define $\sin^2x = t$, the QHJ equation becomes
\begin{equation}
\tilde{p}^2(t)-2i\sqrt{t(1-t)}\frac{d}{dt}\tilde{p}(t) =
E-\frac{A}{t}-\frac{B}{1-t}+Ct-Dt^2    \label{e39}
\end{equation}
 where $\tilde{p}(t) \equiv p(x)$.  Defining $q$ by $\tilde{p}=
\sqrt{t(1-t)}q$ we obtain the QHJ equation in $t$ variable as follows,
\begin{equation}
q^2- 2i\frac{d}{dt}q+\frac{i(1-2t)q}{t(1-t)} =
\frac{E}{t(1-t)}-\frac{A}{t^2(1-t)}-\frac{B}{t(1-t)^2}+\frac{C}{1-t}-\frac{Dt}{1-t}.
\label{e40}
\end{equation}
The quantization condition \eqref{e38}{} becomes
\begin{equation}
\frac{1}{2\pi}\oint_{C} \frac{q(t)}{2}dt = n.   \label{e41}
\end{equation}
We observe that QMF has fixed poles at $ t = 1,0$ and proceeding in a way
similar to \S 4, we obtain the
quasi-exact solvability condition as
\begin{equation}
M=n.   \label{e42}
\end{equation}

\section{Hyperbolic Potential}

The QHJ equation for this potential is
\begin{equation}
p^2(x)-i \frac{d}{dx}p(x) =
E+\frac{A}{\cosh^2x}-\frac{B}{\sinh^2x}+C\cosh^2x-D\cosh^4x   \label{e43}
\end{equation}
where
\begin{equation}
A=4(S_1-\frac{1}{4})(S_{1}-\frac{3}{4})
,\,\,B=4(S_2-\frac{1}{4})(S_{2}-\frac{3}{4})  \label{e44}
\end{equation}
\begin{equation}
C=q_{1}^{2}+4q_1(S_1+S_2+M) ,\,\,D=q_{1}^{2}  \label{e45}
\end{equation}
and the ranges of $S_1$ and $S_2$ are $2S_1>1$ and $2S_2>1$. A change of
variable from $x \rightarrow t = \cosh x$ changes $p(x) \rightarrow
\tilde{p}(t) = p(\cosh^{-1}x)$, then defining
\begin{equation}
\tilde{p}(t)= \sqrt{t^2-1}q(t).   \label{e46}
\end{equation}
The QHJ equation becomes
\begin{equation}
q^2-i \frac{d}{dt}q-\frac{i tq}{t^2-1}=
\frac{E}{t^2-1}+\frac{A}{t^2(t^2-1)}-\frac{B}{(t^{2}-1)^{2}}+\frac{Ct^2}{t^2-1}-\frac{q_{1}^{2}t^{4}}{t^2-1}
   \label{e47}
\end{equation}
The quantization condition is
\begin{equation}
\frac{1}{2\pi}\oint q(t)dt = n     \label{e48}
\end{equation}
We expect 2$n$ moving poles on the entire real line as in Sec. IV, and
fixed poles at $ t=0,\pm1$. These poles together with a
pole of finite order at infinity lead to the condition
\begin{equation}
 M=n.    \label{e49}
\end{equation}
which agrees with known condition of QES of this model.

\section{Conclusions}
In the limit $\hbar \rightarrow 0$ QMF of the sextic oscillator goes to
$p_{cl}$ which has six branch points. Therefore, in general we expect a
complicated singularity structure for the sextic and also for other
models studied in this paper. In the analysis presented in the previous
sections it was assumed that
there are no moving poles off the real axis. A closer look at the
derivation of the condition of quasi-exact solvability shows that the
assumption`` no moving poles of the QMF off the real axis '' can be
replaced by a weaker assumption, `` QMF has a finite number of moving
poles in the complex plane '' without altering any of the
results. In fact it can be seen from explicit solutions  in [2,4] that the
algebraic eigen functions do have complex zeros. For all the models
that have been studied here, the algebraic part of the spectrum and
eigenfunctions are well known and in all the cases QMF has a pole at
infinity. Thus we conclude that, for the class of potentials studied
here, QES models are the only models for which QMF has a pole at
infinity and a finite number of moving poles in the complex plane. For
each model the known algebraic eigenfunctions correspond to QMF having
singularity structure as postulated. For completeness it must be
mentioned that the assumption of finite number of poles is not
independent of the assumption that the point at infinity is a
pole. For a large class of potentials which are analytic everywhere,
except for isolated singularities, the moving singularities of
solutions of QHJ can only be poles. An infinite number of such poles
will, therefore, have an accumulation point at infinity and making it
$z=\infty$
a non-isolated essential singular point.\\ The
parameter $n$ that appears in the exact quantization condition is
related to the number of moving poles in QMF and has different roles
to play for ESM and QES models. In ESM each value of n corresponds to an
energy level and an eigenfunction with $n$ real zeros. In case of the
QES models, $n$ appears as a parameter in the expression for the potential
and picks out a particular QES model within a family of potentials ;
varying $n$ gives rise to a different potential within the
family. Recalling that $n$ also determines the number of moving poles
of the QMF it appears reasonable to expect that all the algebraic
eigenfunctions for a given QES models (fixed $n$) will have the same
number
of the complex zeros determined by $n$. An explicit check reveals
that this expectation is true for the sextic oscillator. A preliminary
study reveals that,
the poles of the QMF for QES periodic potentials have a richer structure.
A detailed study
of the location of the poles of the QMF in different QES models will be
reported elsewhere. \\

{\bf Acknowledgments}
The authors acknowledge useful discussions with \mbox{P.K. Panigrahi}. KGG
would like to  thank UGC for the financial support. AKK
acknowledges useful discussions with S. Chaturvedi and V. Srinivasan.

%\newpage

{\bf References}
\begin{enumerate}

\item Singh V, Biswas S N, Datta K, 1978 {\it Phys. Rev} {\bf D18} 1901

\item  Ushveridze A 1994 {\it Quasi-exactly Solvable Models in
    Quantum Mechanics} (Bristol: Institute of Physics Publishing)

\item  Gon$\acute{z}$alez-L$\acute{o}$pez A, Kamran N and Olver P J 1993
{\it Commun. Math. Phys} {\bf 153} 117

\item  Gon$\acute{z}$alez-L$\acute{o}$pez A, Kamran N and Olver P J 1994
{\it Contemp. math} {\bf 160} 113

\item Bagchi B, Mallik S, Qussne C and Roychoudhary R (2001) {\it Phys
Lett A } {\bf 289}

\item khare A and Mandal B P (2000) {\it Phys Lett A} {\bf 272}

\item Dunne G V and Shiffman M (2002){\it Annals Phys} {\bf 299} 143

\item Khare A (2001) {\it Phys Lett A} {\bf 288} 69

\item Thatuk V M and Voznyak O  (2002) {\it Phys Lett A} {\bf 301} 177

\item Bender C M and Boettcher S (1998) {\it Jour Phys A } Math and Gen
{\bf 31} L273

\item Turbiner A V and Ushveridze A G (1987) {\it Phys Lett A } {\bf 126}
No 3 181

\item Dunne G and Mannix J (1998) {\it Phys Lett B} {\bf 428} 115

\item   Panigrahi P K and Atre R, (under preparation) {\it Quasi-Exactly
Solvable Problems : A New
     Approach and an approximation Scheme}
\item  Leacock R A and Padgett M J, 1983 {\it Phys. Rev. Lett}. {\bf 50} 3

\item  Leacock R A and Padgett M J, 1983 {\it Phys. Rev.} {\bf D28} 2491

\item  Bhalla R S, Kapoor A K  and Panigrahi P K 1997 {\it Am. J. Phys.}
{\bf 65} 1187 (1997).

\item  Bhalla R S, Kapoor A K  and Panigrahi P K 1997 {\it Mod. Phys.
Lett.} {\bf A12} 295

\item Jayanthi S 1998 {\it Study Of One Dimensional Potential Problems
using Quantum Hamilton Jacobi Formalism}, M.Phil thesis submitted to the
    University of Hyderabad

\item  Hautot A P 1972 {\it Phys. Lett.} {\bf 38A} 305

\end{enumerate}
\end{document}